\documentclass[preprint,12pt,authoryear]{elsarticle}
\usepackage[left=1in,right=1in,top=1in,paper=a4paper]{geometry}

\usepackage{amssymb}
\usepackage{amsmath}
\usepackage{amsfonts}
\usepackage[colorlinks=true,linkcolor=black,anchorcolor=black,citecolor=black,filecolor=black,menucolor=black,runcolor=black,urlcolor=black]{hyperref}
\usepackage{url} 
\usepackage{lineno}
\usepackage{mathtools}
\usepackage{nicefrac}
\usepackage[noend]{algpseudocode}
\usepackage{algorithm}
\usepackage{soul}
\usepackage{titlesec}
\usepackage{rotating}
\usepackage{svg}
\usepackage[skip=6pt]{caption}
\DeclareMathOperator*{\argmin}{argmin}
\titlespacing*{\subsection}{0pt}{\baselineskip}{\baselineskip}

\journal{Advances in Water Resources}

\begin{document}

\begin{frontmatter}



\title{Dynamic Mode Decomposition of real-time 4D imaging data to explore intermittent fluid connectivity in subsurface flows}

\author[Stanford]{Aman Raizada}
\author[Shell]{Steffen Berg}
\author[Stanford]{Sally M. Benson}
\author[Stanford]{Hamdi A. Tchelepi}
\author[Stanford]{Catherine Spurin}

\affiliation[Stanford]{organization={Department of Energy Science and Engineering},
            addressline={Stanford University}, 
            city={Stanford},
            state={CA},
            country={USA}}

\affiliation[Shell]{organization={Shell Global Solutions International B.V.},
            addressline={2288 GS Rijswijk}, 
            country={The Netherlands}}

\begin{abstract}
The interaction of multiple fluids through a heterogeneous pore space leads to complex pore-scale flow dynamics, such as intermittent pathway flow. The non-local nature of these dynamics, and the size of the 4D datasets acquired to capture them, presents challenges in identifying key fluctuations controlling fluid connectivity. To address these challenges, this work employs Dynamic Mode Decomposition (DMD), a data-driven algorithm that decomposes complex nonlinear systems into dominant spatio-temporal structures without relying on prior system assumptions.

We present a workflow that identifies critical spatio-temporal regions exhibiting intermittent flow dynamics. This workflow is validated through three test cases, each exploring the impact of viscosity ratio on flow dynamics while maintaining a constant capillary number. Our findings demonstrate DMD's potential in analyzing extensive experimental datasets and identifying crucial intermittent flow structures, offering a powerful tool for understanding complex fluid behaviors in heterogeneous pore spaces. Using our method, we can quickly identify the timescales and locations of interest in an objective manner, providing a valuable diagnostic tool for analysing large synchrotron datasets.

\end{abstract}

\begin{highlights}
\item Dynamic Mode Decomposition is used to locate dynamics in large 3D datasets.
\item We observe how dynamics evolve as the viscosity ratio is altered.
\end{highlights}


\end{frontmatter}


\section{Introduction}

Multi-phase fluid flow in the subsurface is commonly modeled with a phenomenological extension of Darcy's law \citep{bear2012introduction,darcy1856fontaines}. The inherent assumption in this model is that each fluid occupies its own static pathway, connected across the pore space \citep{blunt2017multiphase,dullien2012porous}. However, high-resolution, fast X-ray imaging has shown dynamic fluid interfaces that depend on the complex topology of the pore space and fluids within \citep{reynolds2017dynamic, rucker2015connected, spurin2019intermittent,spurin2020real,gao2019pore,gao2020pore}. The onset of dynamics in a given system depends on the capillary and viscous forces \citep{spurin2019mechanisms}.

Intermittent flow pathways, where fluid flow pathways periodically connect and disconnect, have been observed even while macroscopic fluid flow properties, such as saturation are constant, when averaged over time \citep{reynolds2017dynamic,spurin2019intermittent, spurin2020real}. The complex interaction between a heterogeneous porous rock and the interface among percolating fluids plays an important role in the potential development of such unique flow patterns. This interaction can result in the creation of localized favorable flow paths. Consequently, the nonlinear dynamic behavior can lead to the entrapment or remobilization of nonwetting phase clusters at the pore-scale. It is important to consider these physics for accurately modeling CO$_2$ trapping mechanisms, such as residual trapping \citep{ni2019predicting}, to ensure secure and long-term geologic carbon storage. There exists a rich body of literature describing macroscale forward-models based on the method of volume averaging \citep{bertin1990two,crapiste1986general,wood2013volume}, thermodynamically constrained averaging theory \citep{gray2005thermodynamically,hassanizadeh1993thermodynamic,gray2014introduction}, homogenization theory \citep{hornung1997homogenization,auriault1995taylor}, and others (see \citep{battiato2019theory} and the references therein). However, mechanistic models that capture complex pore-scale dynamic phenomena such as intermittent pathway flow, are currently lacking.

The intermittent fluctuations have been observed over a wide range of rock types, at different length scales. However, large 4D synchrotron experimental datasets coupled with complex non-local flow dynamics leads to difficulties in locating key fluctuations that control fluid connectivity.  Identifying intermittent pathways in a heterogeneous pore-space, and inferring their temporal dynamics by sifting through a stack of images proves extremely challenging and time-intensive when relying solely on the naked eye. These challenges motivate us to explore an analytical approach which is capable of effectively and computationally efficiently extracting important dynamic features from the flow data.

In this study, we use a data-driven decomposition algorithm called Dynamic Mode Decomposition (DMD) to explore fluctuations in the pore-space \citep{schmid2022dynamic, taira2017modal}. DMD has the following important characteristics: (1) it relies purely on input data via a sequence of snapshots, (2) it employs an equation-free architecture, (3) it avoids making any prior assumptions about the system, and (4) it provides an accurate decomposition of a complex nonlinear system into dominant spatio-temporal coherent structures and patterns \citep{kutz2016dynamic}. Overall, the described characteristics make DMD a suitable choice for analyzing our extensive experimental datasets because: (1) a predictive model based on governing equations is unknown in this case; (2) empirical knowledge indicates that the system exhibits idiosyncratic behavior related to dynamic connectivity \citep{reynolds2017dynamic,spurin2019intermittent}; and (3) extracting dominant structures can help in identifying the critical locations in the pore space that exhibit intermittency.

We build upon the conclusions presented in \citep{spurin2023dynamic}, which demonstrate that DMD can effectively decompose a 4D signal and reconstruct the input data using a limited number of mode-dynamic pairs. However, the analysis in \citep{spurin2023dynamic} lacked an algorithm for systematically and effectively identifying the meaningful dominant coherent structures from the full suite of modes resulting from the data decomposition. To address this limitation, in this study, we develop a novel workflow that separates the convolved spatio-temporal space of experimental data into distinct spatial regions exhibiting intermittent behavior that evolves over both short- and long-time scales. We use DMD to identify key locations within the heterogeneous pore space that exhibit intermittent flow behavior. For the training data, we examine the saturation field in a flow regime where the viscosity ratio (M), previously identified as a key parameter controlling the development of intermittency \citep{spurin2019mechanisms}, is varied while keeping the capillary number ($Ca$) constant in the $Ca$-M parameter space. Based on the workflow, three unique scenarios are presented, which demonstrate the excellent accuracy of the DMD results.


\section{Data and Methods} \label{method}

\subsection{Experimental Data Acquisition}

\begin{figure}[!t]
\centering
\includegraphics[width=12.9cm]{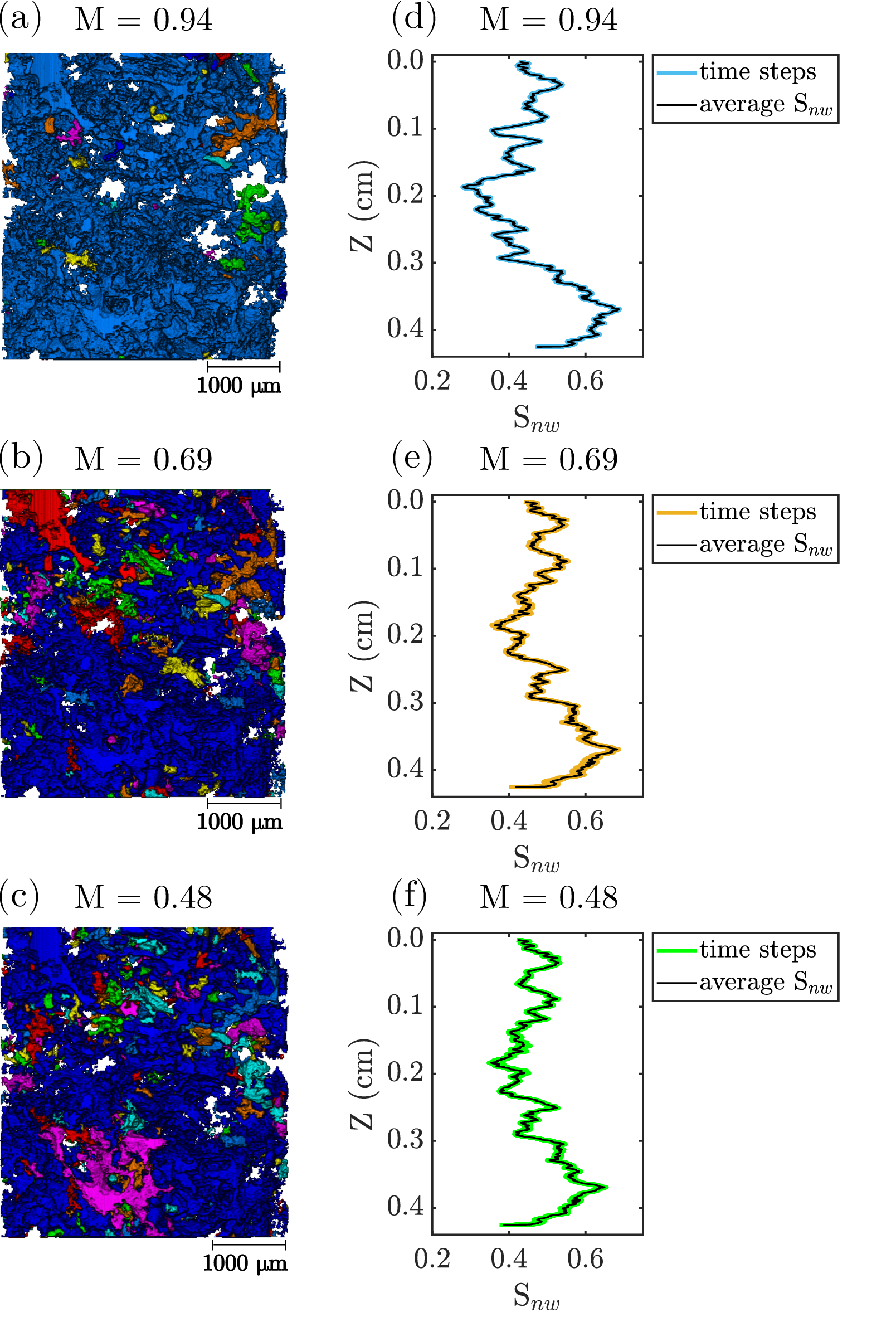}
\caption{The disconnected ganglia of the nonwetting phase are shown in different colors for: (a) M = 0.94, (b) M = 0.69, and (c) M = 0.48, i.e., each connected ganglion is represented by a single color. The number of ganglion increases as the viscosity ratio decreases. The fluid flow occurs from the bottom toward the top of the sample. In addition, panels (d) M = 0.94, (e) M = 0.69, and (f) M = 0.48, show the 2D slice-averaged saturation S$_{nw}$ for all time steps along the axial direction of the sample. The black curve in figures (d), (e), and (f) represents the temporally averaged S$_{nw}$ value at each slice position along the length of the sample.}
\label{gang_avg_map}
\end{figure}

The experiments were conducted in a cylindrical Estaillades carbonate sample with a diameter of 5~mm and a length of 21~mm. The sample was attached to a dual injection piece, specifically designed for the simultaneous injection of two fluid phases while suppressing the generation of slug flow. The sample was initially saturated with brine (deionized water doped with 15\% wt. KI to improve the X-ray contrast). The system was pressurized to 8~MPa, with an additional confining pressure of 2~MPa. Next, both decane and brine were injected simultaneously at a fractional flow ($f_w$) of 0.5. The steady-state (or statistically stationary) regime was determined to be reached when macroscopic variables, such as saturations and differential pressure across the rock core, began to fluctuate around a constant value \citep{tallakstad2009steady,tallakstad2009steady1,gao2017x,spurin2019intermittent}. Once the system reached a steady-state (determined by the pressure drop across the sample), for a given viscosity ratio M (M $=\nicefrac{\mu_{nw}}{\mu_w}$), 3D snapshots of the pore space were captured using synchrotron tomography. These snapshots were taken at time intervals between t = 0 seconds, which is the initial image, and 98 seconds. The viscosity ratio was varied during the experiments by introducing glycerol into the brine. The initial experiment, which corresponds to M = 0.94, did not have any glycerol mixed with brine. For the M = 0.69 and M = 0.48 experiments, glycerol was mixed with brine in the ratio 1:9 and 2:8, respectively. Note that the sample was not resaturated with brine between the different experiments. The variation in M allowed for changes in the total flow rate while keeping the total capillary number constant ($Ca \approx 1.6\times 10^{-6}$) for all cases.

The images captured at the synchrotron have an isotropic voxel size of 2.75 \textmu m. The core was only imaged near the middle region and the length of this investigated region was approximately equal to 0.43 cm. This length corresponds to a total of 1550 two-dimensional slices of saturation data spread uniformly along the axial direction of the sample. A single scan, or in other words, a time snapshot, was acquired over a period of 1 second, while the time difference between successive scans was 2 seconds. In total, the data repository for the cases where M = 0.48 and M = 0.69 consists of 50 time snapshots, while there are 49 snapshots for the M = 0.94 experiment. Thus, the initial time snapshot and subsequent data capture the simultaneous two-phase flow behavior at the pore-scale under dynamic flow conditions in a steady-state regime. 


Figures \ref{gang_avg_map}(a)-(c) show the disconnected ganglia of the nonwetting phase (decane) in different colors at the first time step for all experiments. Here, each single connected ganglion is represented in a distinct color. The fluid flow direction is from the bottom toward the top of the sample. As the viscosity ratio decreases in Figures \ref{gang_avg_map}(a) to (c), more pore space is occupied by the nonwetting phase and a higher number of disconnected ganglia are observed. Overall, the network of ganglia appears to be macroscopically static over time; however, there is a dynamic evolution in the count of ganglia. This is primarily due to the active rearrangement of fluid interfaces, which occurs because of flow instabilities at critical locations within the pore space, even though the volumetric saturation of the sample remains relatively static. Identifying such critical locations in the pore space with the naked eye is notoriously difficult. Also, in order to capture these dynamically evolving fluid menisci configurations, the temporal resolution of images must be sufficiently high. The time required to create fluid interfaces under steady-state conditions in capillary-dominated regimes can be estimated by $t=\nicefrac{{\sigma K_{eff}}}{{l \mu q_t^2}}$ \citep{reynolds2017dynamic}, where $\sigma$ is the interfacial tension between the nonwetting and wetting phases, $K_{eff}$ is the effective permeability in multi-phase flow, $q_t$ is the total Darcy velocity, and $l$ is the characteristic pore size. For an order of magnitude estimate corresponding to the lowest viscosity ratio experiment, we consider $K_{eff}=1.06\times 10^{-13}$ m$^2$ (from \citep{gao2019pore}), $l=19.58\times 10^{-6}$ m, average viscosity of fluids $\mu=82.9\times 10^{-5}$ Pa$\cdot$s, $\sigma=64 \times10^{-3}$ N/m, and $q_t=2.34\times 10^{-5}$ m$\cdot$s$^{-1}$ (remaining parameters from \citep{spurin2021intermittent}). The timescale for fluid interface creation was calculated to be approximately 13 minutes. Therefore, the time interval between successive scans (2 seconds) is sufficient to capture any alterations in fluid configurations in the core sample caused by external flow energy provided through injection. 

Figures \ref{gang_avg_map}(d)-(f) display the 2D slice-averaged S$_{nw}$ for all time steps corresponding to different viscosity ratios. The black curve in these plots shows the temporally averaged S$_{nw}$ value at each slice position along the length of the sample. The spread of S$_{nw}$ time step curves around the mean S$_{nw}$ trend is largest for the M = 0.48 case (green region in Figure \ref{gang_avg_map}(f)) and smallest for the M = 0.94 experiment (blue region in Figure \ref{gang_avg_map}(d)). This observation indicates that as the viscosity ratio decreases (while keeping other parameters constant), there is a growing influence of localized heterogeneity on fluid flow pathways across the sample, leading to increased fluctuations in S$_{nw}$. 

\begin{figure}[!b]
\centering
\includegraphics[width=16cm]{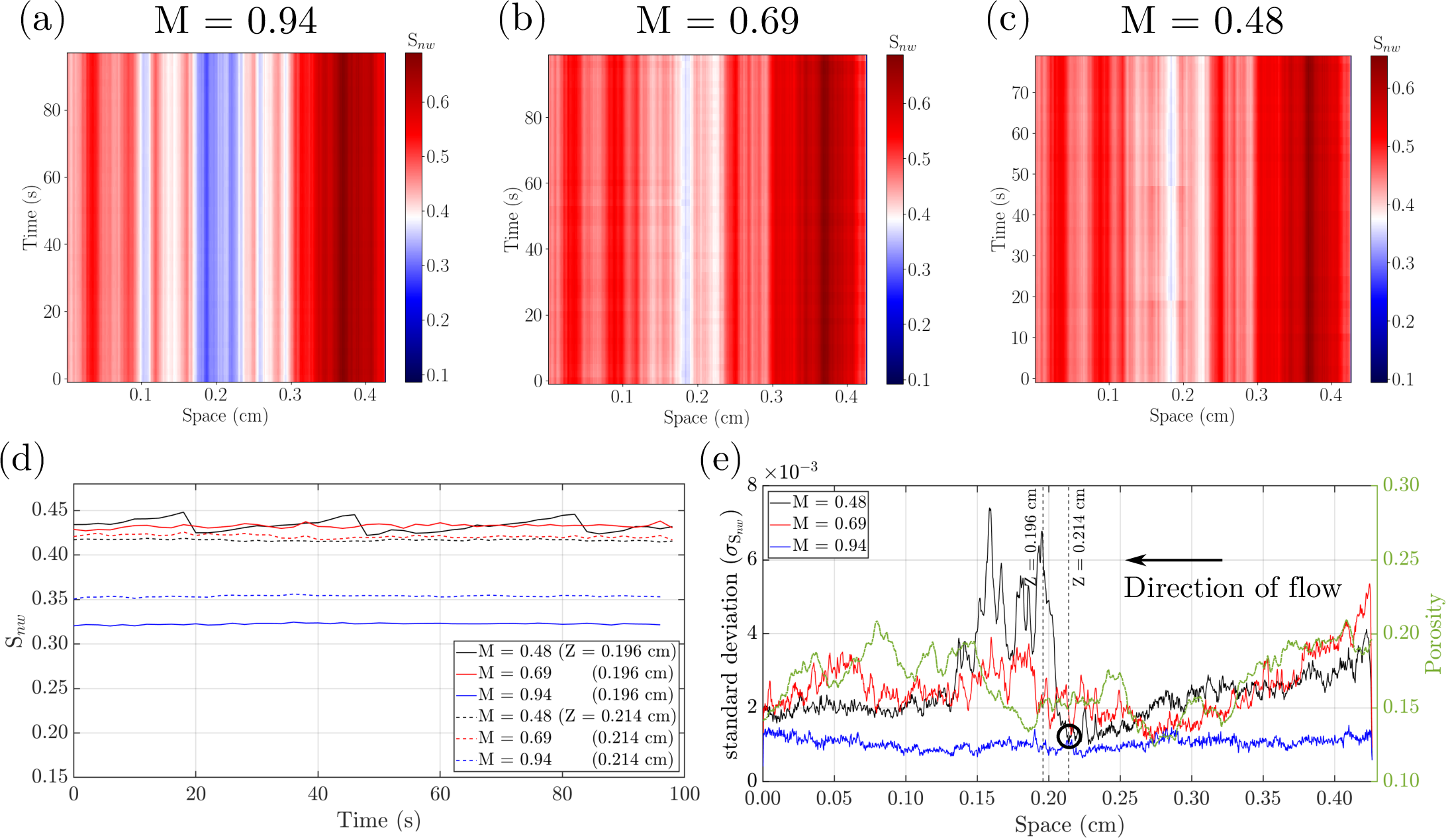}
\caption{The spatio-temporal maps of S$_{nw}$ are shown for: (a) M = 0.94, (b) M = 0.69, and (c) M = 0.48. (d) Temporal evolution of S$_{nw}$ at two locations in the sample, Z = 0.196 cm and 0.214 cm, for different M, and (f) standard deviation in S$_{nw}$ across time along the length of the sample for different M. In figures (a)-(c) and (e), the direction of flow is from right to left.}
\label{sat2locs}
\end{figure}

The spatio-temporal evolution of S$_{nw}$ is displayed across the 1D space and time for different viscosity ratio cases in Figures \ref{sat2locs}(a)-(c). Here, the influence of localized pore heterogeneity is evident in the M $=0.94$ case, where low S$_{nw}$ can be observed near the middle region of the sample when compared to the other viscosity ratios. To illustrate the role of time dynamics in the experiments, Figure \ref{sat2locs}(d) shows the temporal evolution of slice-averaged S$_{nw}$ for two arbitrarily chosen positions along the sample, at Z = 0.196 cm and Z = 0.214 cm. At Z = 0.196 cm, there are large variations in S$_{nw}$ for the M = 0.48 case, and this reflects to some extent as a periodic behaviour of the system. Comparatively, there is less variation observed in S$_{nw}$ for the M = 0.69 case, and it is least for the M = 0.94 experiment. Similar inferences can be drawn from the standard deviation in S$_{nw}$, calculated across all time snapshots and shown in Figure \ref{sat2locs}(e). At Z = 0.196 cm in Figure \ref{sat2locs}(e), the second global maximum occurs for M = 0.48, whereas the standard deviation for M = 0.94 remains the smallest. Another important observation can be made at Z $\sim$ 0.214 cm in Figure \ref{sat2locs}(e). Here, it is evident that the standard deviations for all viscosity ratio cases tend to reduce in the vicinity of this region and eventually converge to a common value. The porosity at this location is approximately 0.15, as shown on the right y-axis in Figure \ref{sat2locs}(e). This is significantly different from the global minimum value of 0.123 at Z = 0.274 cm. However, the $\sigma_{\mathrm{S}{_{nw}}}$ values for all M are nearly equal. This observation further emphasizes that, while significant pore space exists in this region, it is the poor spatial connectivity within the pore structure that hinders fluid flow. Overall, the maximum changes in $\sigma_{\mathrm{S}{_{nw}}}$ are observed in the case of M = 0.48, whereas they are least pronounced in the M = 0.94 case. This suggests that as the viscosity ratio is reduced from 0.94 to 0.48, the multi-phase flow system exhibits significant nonlinear dynamical behavior.



\subsection{Dynamic Mode Decomposition}
The dynamic mode decomposition is a matrix factorization technique that decomposes time-resolved data into spatial modes representing dominant coherent structures in the data, sharing the same frequency or wavenumber. The time dynamics associated with the DMD modes are constrained to be linear. This implies that each spatial mode is linked to a complex frequency, indicating an oscillatory behavior, potentially with an exponential growth or damping rate. A major advantage of DMD, compared to other popular dimensionality reduction techniques such as proper orthogonal decomposition (POD) \citep{liang2002proper}, is that the spatial modes obtained from DMD exhibit temporal orthogonality and, in general, are spatially non-orthogonal \citep{schmid2011applications}. This means that it is possible to extract the temporal evolution (pure frequencies) of the modes due to the constraints imposed by the linearity of the time dynamics. A simple description of the DMD method is outlined below.

Consider a sequence of experimental or high-fidelity simulation measurements (column vectors) \{$x_1, ..., x_m$\}, where $m$ denotes the number of snapshots constituting the data sequence and $x_{k} \in \mathbb{R}^{n}$ represents an $n$-dimensional state vector at time step $t_k$. The spatial dimension $n$ of a snapshot is typically much greater than the number of snapshots $m$, i.e., $n \gg m$. In this study, $m$ = 49 for the M = 0.94 experiment and $m$ = 50 for M = 0.69 and 0.48 cases. The spatial dimension $n$ of each column vector is $1550\times 1$, corresponding to the 2D slice-averaged S$_{nw}$ data points along the axial direction of the sample for all viscosity ratios. Next, the sequence of snapshots are arranged into two matrices of the form:\\

\textbf{X} = $\begin{bmatrix}
 | & | &   & | &  & |\\
 x_1 & x_2 & ... & x_k & ... & x_{m-1} \\
 | & | &   & | &  & |
\end{bmatrix}$, \quad \textbf{Y} = $\begin{bmatrix}
 | & | &   & | &  & |\\
 x_2 & x_3 & ... & x_{k+1} & ... & x_{m} \\
 | & | &   & | &  & |
\end{bmatrix}$.
\\ \\

The DMD algorithm finds the best-fit matrix \textit{A} capable of approximating the simultaneous evolution of measurements from the state \textbf{X} to \textbf{Y}, i.e., one time step into the future, such that 

\begin{equation} \label{eq1}
    \textbf{Y}\approx \textit{A}\textbf{X} \,.
\end{equation}

The linear mapping operator \textit{A} in Equation \ref{eq1} can be estimated by solving the following optimization problem:

\begin{equation}
    \textit{A} = \argmin \limits_{\textit{A}} ||\textbf{Y}-\textit{A}\textbf{X}||_F = \bold{Y}\bold{X^+}\,,
\end{equation} where $||\textbf{M}||_F := \sqrt{\Sigma_{i}\Sigma_{j} m^2_{i,j}}$ is the Frobenius norm and $\bold{X^+}$ is the Moore-Penrose pseudoinverse \citep{moore1920reciprocal} of the non-square matrix $\textbf{X}$. Further, the DMD uses singular value decomposition (SVD) to produce a rank-reduced representation of matrix \textit{A} by projecting it onto a reduced-dimensional subspace defined by $r$ POD modes. Next, the spectral decomposition of this reduced-rank matrix $\tilde{A}$ is computed as $\tilde{A}\,\mathbf{W}= \boldsymbol{\Lambda} \mathbf{W}$. The eigenvalues $\lambda_j$ of $\tilde{A}$ are contained in the diagonal matrix $\boldsymbol{\Lambda}$, whereas the columns of matrix $\mathbf{W}$ represent the eigenvectors of $\tilde{A}$. Further, the eigenvectors $\phi_j$ of matrix \textit{A} can be computed via a linear transformation of dynamic modes $\mathbf{W}$ corresponding to $\tilde{A}$. (For more details about the DMD algorithm, the reader is referred to \citep{tu2013dynamic}.) Finally, the reconstructed data at time $k\Delta t$ can be obtained from the linear combination of DMD modes as follows:

\begin{equation} \label{eq3}
    x_{k} \approx \displaystyle\sum\limits_{j=1}^{r} \phi_j \lambda_j^{k-1} b_j = \mathbf{\Phi} \mathbf{\Lambda}^{k-1} \mathbf{B} \,,
\end{equation}
where $\mathbf{B}$ is a diagonal matrix containing amplitudes $b_j$'s of the DMD modes, the columns of $\mathbf{\Phi}$ represent the eigenvectors of \textit{A}, and the eigenvalues $\mathbf{\Lambda}$ with $\lambda_i \in \mathbb{C}$ contain information about the growth/decay rates and frequencies of the DMD modes. 

In addition, we use two performance metrics to evaluate the accuracy of the models reconstructed using the DMD method. First, the relative reconstruction loss ($\mathcal{L}$) is defined as:

\begin{equation}
    \mathcal{L} = \frac{||{\mathbf{Z}^{true}}-{\hat{\mathbf{Z}}}||_F}{{||\mathbf{Z}^{true}}||_F}\,,
\end{equation}
where $\mathbf{Z}^{true}$ represents the true data matrix and ${\hat{\mathbf{Z}}}$ refers to the matrix reconstructed using DMD based on the reduced rank $r$. Second, we calculate the Mean Absolute Error (MAE) as follows:

\begin{equation}
    \mathrm{MAE} = \frac{1}{m\,n} \,{\displaystyle\sum\limits_{i=1}^{m} \displaystyle\sum\limits_{j=1}^{n}|\mathbf{Z}_{ij}^{true}-\hat{\mathbf{Z}}_{ij}}|\,,
\end{equation} 
to evaluate the average difference between the original and DMD reconstructed datasets. A decrease in the value of aforementioned performance metrics indicates an increase in the accuracy of the DMD method in reconstructing the experimental datasets using limited ranks.   

\section{Results}

\subsection{Singular value decomposition (SVD) for dimensionality reduction of experimental data} 
\label{res1}

\begin{figure}[!t]
\centering
\includegraphics[width=15cm]{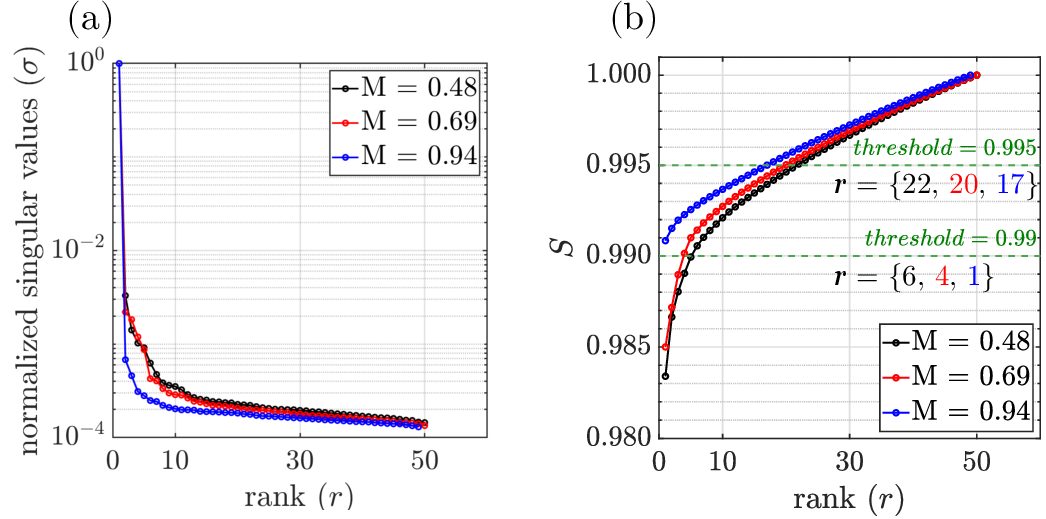}
\caption{The trend of (a) normalized singular values and (b) the normalized cumulative sum ($S$) of the first $r$ singular values in the singular value decomposition (SVD) of S$_{nw}$ for different viscosity ratios.}
\label{SVD}
\end{figure}

We present the SVD results to gain insights into the structure and dimensionality of our data. In addition, we explore whether the data can be effectively represented with a sufficiently low number of ranks. For this analysis, we examine the distribution of the singular values obtained from SVD and compute their cumulative sum ($S$) as follows:
\begin{equation}
    S = \frac{\Sigma_{i=1}^{r} \sigma_i}{\Sigma_{i=1}^{m} \sigma_i} \,,
\end{equation}
which represents the ratio of the total variance in the dataset contained in the first $r$ singular values. Figures \ref{SVD}(a) and \ref{SVD}(b) display the trends in the normalized distribution of singular values and $S$, respectively, with respect to the rank number. From these plots, we can identify the importance of modes and the extent to which retaining the first $r$ ranks captures the variance within the dataset. The maximum number of ranks is determined by the total number of time snapshots ($m$). For the M = 0.94 case, $m$ is equal to 49, while for M = 0.69 and 0.48 experiments, $m$ is equal to 50.

The distribution of the normalized singular values, as shown in Figure \ref{SVD}(a), indicates that a significantly large proportion of the system's energy is carried by the first dominant value. This value is at least two orders of magnitude greater than the succeeding singular value for all viscosity ratios. This numerical finding is in agreement with the physical behavior observed in the oil-brine experimental system, where it is noted that a large network of established connected flow pathways mostly remains unchanged over time. As a result, the volumetric saturation of the sample remains temporally stable, with minimal fluctuations observed in the mean behavior of the system. This stability is manifested in the first (and most dominant) singular value in SVD across all viscosity ratios.

To quantitatively estimate the minimum SVD rank required for an accurate reconstruction of the original data, we choose to determine the smallest value of $r$ that sufficiently captures at least 99\% of the total variance in the data, i.e., $S \geq 0.99$. The dashed line in Figure \ref{SVD}(b) intersects the curves for different M, allowing us to estimate the number of necessary ranks determined by this criterion. For the M = 0.94 case, only the first rank is necessary. In the cases of M = 0.69 and M = 0.48, 4 and 6 ranks are required, respectively, to capture at least 99\% energy of the system. The SVD result is qualitatively consistent with the multi-phase flow behavior of the system, where the contribution of intermittency (or flow instability) to the transportation of nonwetting phase across the sample is expected to increase with a decrease in viscosity ratio \citep{spurin2019mechanisms}. Thus, as the viscosity ratio decreases, a greater number of ranks are required to capture the emergence of intermittent flow behavior compared to the mostly connected and stable fluid pathways observed in the highest viscosity ratio case. We also observe that on increasing the threshold of $S$ to 0.995, there is a notable increase in the number of ranks needed to capture a larger proportion of the total variance in the data (for the cases M = 0.94, 0.69, and 0.48, the values of 
$r$ are 17, 20, and 22, respectively, as shown in Figure \ref{SVD}(b)).

The number of ranks correlates with the amount of intermittent pathway flow. We explore this further by using the values of $r$ corresponding to the scenario where $S \geq 0.99$ for different M. We refrain from considering a higher number of ranks in correspondence to the higher threshold case for the following reasons: (1) We use 2D slice-averaged saturation, which may smooth out small-scale local fluctuations in the nonwetting phase volume, manifesting in finer variations in saturation on a slice and posing a challenge for detection; and (2) our spatio-temporal datasets are limited, potentially leading to errors due to uncertainty and a lack of sufficient training data to capture intermittent behavior at a smaller length scale.

\subsection{Data reconstruction using DMD} \label{res2}

\begin{figure}[!t]
\centering
\includegraphics[width=16cm]{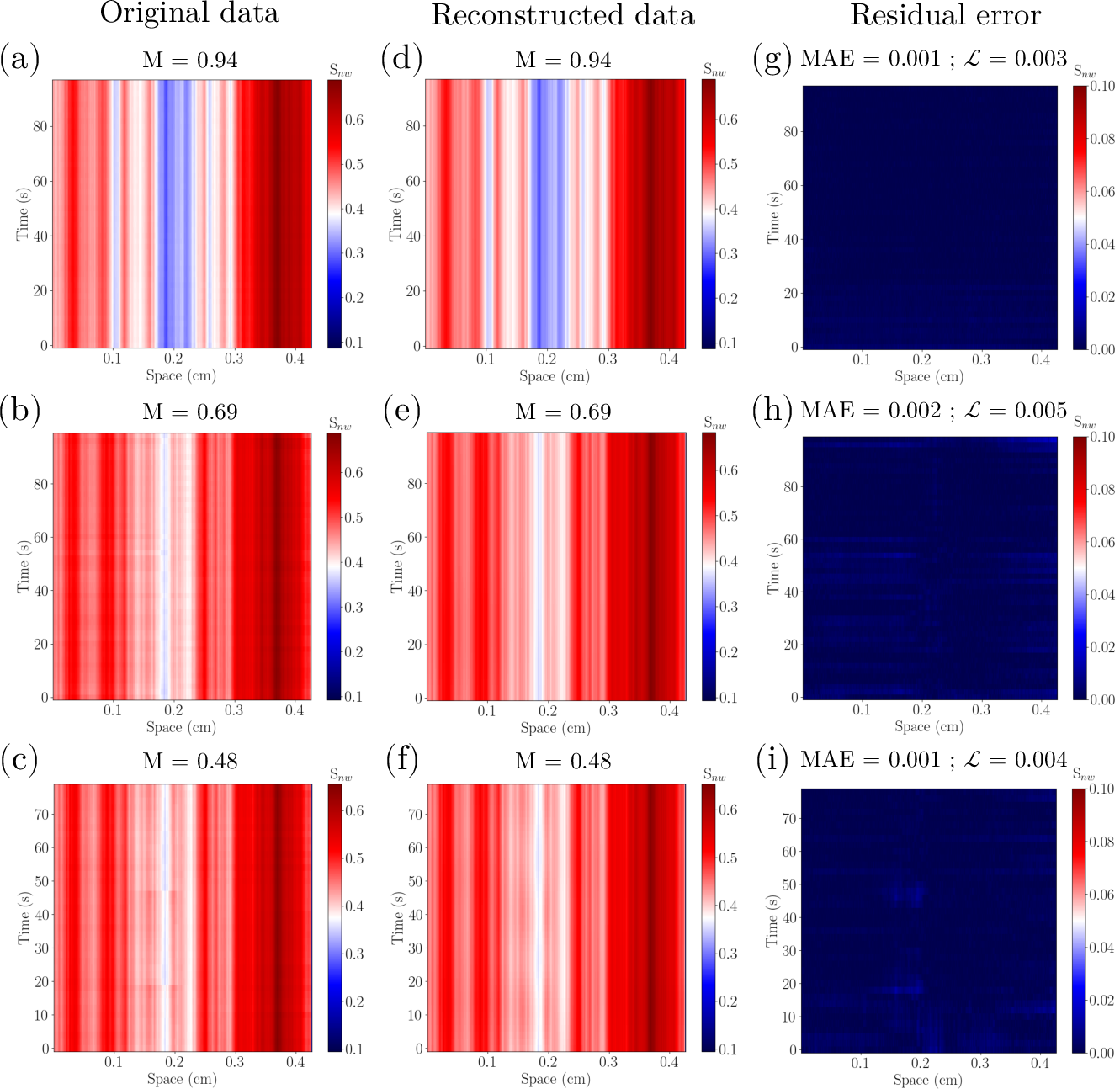}
\caption{The spatio-temporal maps of the original S$_{nw}$ data are shown for: (a) M = 0.94, (b) M = 0.69, and (c) M = 0.48. The S$_{nw}$ data reconstructed using truncated ranks through DMD are shown for: (d) M = 0.94, (e) M = 0.69, and (f) M = 0.48. The residual plots showing the absolute difference in S$_{nw}$ between the original and reconstructed data are displayed in panels (g)-(i) for different M.}
\label{error}
\end{figure}

We compute the DMD of the experimental S$_{nw}$ datasets corresponding to different viscosity ratios. The respective DMD models are then used to reconstruct the data based on truncated ranks, as estimated in the previous section. For M = 0.94, only the first DMD mode is used to reproduce the data, whereas 4 and 6 DMD modes are used for M = 0.69 and 0.48 cases, respectively. Figures \ref{error}(a)-(c) illustrate the spatio-temporal original datasets, while Figures \ref{error}(d)-(f) display the saturation maps reconstructed using truncated DMD modes and their corresponding temporal dynamics. The absolute difference in S$_{nw}$ between the original and reconstructed datasets is shown in the residual error plots in Figures \ref{error}(g)-(i). Overall, the DMD results show a strong qualitative and quantitative agreement with the original datasets. In particular, DMD captures the highly heterogeneous region near the mid-section of the sample, as previously illustrated in Figures \ref{gang_avg_map} and \ref{sat2locs}, where smaller S$_{nw}$ values were noted in the M = 0.94 case.

The influence of intermittency increases with a decrease in viscosity ratio. Thus, in cases with smaller M, where higher levels of intermittency are prevalent due to significant pressure oscillations around the capillary entry pressures, noise generation within the nonlinear dynamical system might become more pronounced \citep{spurin2020real}. This observation is supported by the residual plots in Figures \ref{error}(g)-(i), where both MAE and $\mathcal{L}$ are observed to increase as M decreases. Nonetheless, the maximum MAE and $\mathcal{L}$ are only 0.2\% and 0.5\%, respectively, for the M = 0.69 experiment. This highlights the excellent accuracy achieved in data reconstruction using DMD. Overall, the statistical information of the original 4D signal contained in the truncated ranks has been sufficient for the DMD modes to accurately capture and represent the meaningful patterns in the entire dataset, as depicted in Figures \ref{error}(d)-(f).


\subsection{DMD components} 

\begin{sidewaysfigure}
  \centering
  \includegraphics[width=23cm]{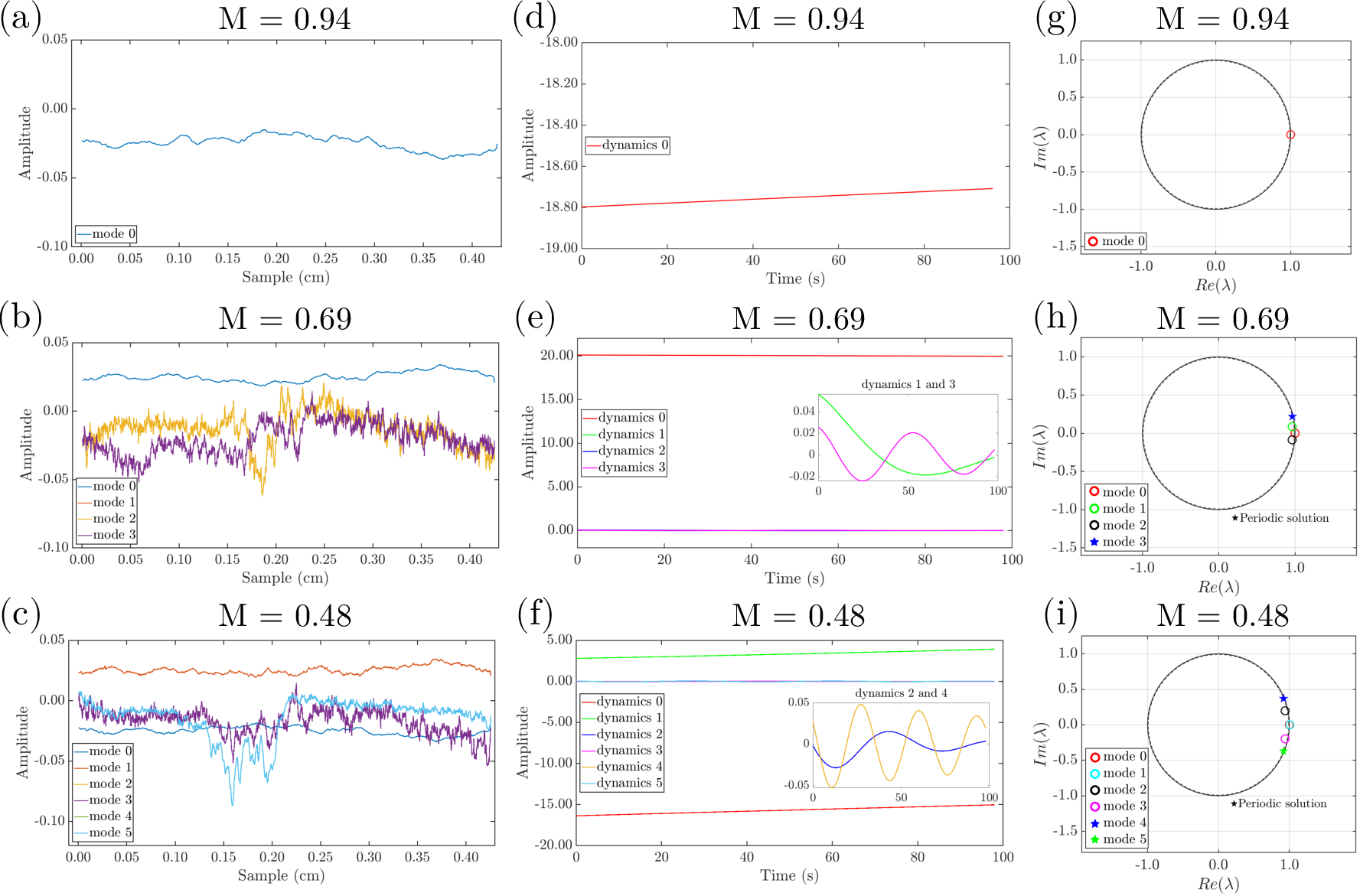}
  \caption{The DMD modes are shown for: (a) M = 0.94, (b) M = 0.69, and (c) M = 0.48. The corresponding time dynamics for these modes are depicted for: (d) M = 0.94, (e) M = 0.69, and (f) M = 0.48. Insets in (e) and (f) highlight the small amplitude of other time dynamics compared to dynamics 0 in both viscosity ratio cases. Finally, the eigenvalues of the DMD modes in the complex plane are shown for: (g) M = 0.94, (h) M = 0.69, and (i) M = 0.48.}
  \label{DMDcomponents}
\end{sidewaysfigure}

DMD identifies prominent coherent structures, often referred to as patterns or modes, within the data \citep{spurin2023dynamic}. Each derived mode is linked to linear dynamics through an eigenvalue, which determines the time evolution of that particular coherent mode. In this section, we illustrate the DMD components for the truncated number of ranks in Figure \ref{DMDcomponents}.  

Figures \ref{DMDcomponents}(a)-(c) display the real part of the spatial modes which correspond to low-dimensional coherent structures present in the saturation data. Figures \ref{DMDcomponents}(d)-(f) display the real part of their temporal evolution for different viscosity ratio experiments. It must be noted that the eigendecomposition of real-valued data using DMD can yield complex eigenvalues and, consequently, complex conjugate pairs \citep{kaptanoglu2020characterizing}. As a result, there is an overlap of curves corresponding to these complex conjugate pairs in Figures \ref{DMDcomponents}(a)-(f). In addition, the eigenvalues associated with the dominant structures are shown in Figures \ref{DMDcomponents}(g)-(i).


We designate the most dominant mode, as determined by the DMD analysis, as mode 0, followed by the subsequent modes in the hierarchy labeled as mode 1, and so forth in Figures \ref{DMDcomponents}(a)-(c). Mode 0, with its linear time dynamics (dynamics 0 in Figures \ref{DMDcomponents}(d)-(f)), signifies the steady-state evolution of the system. This is validated by the eigenvalue being purely real, with $\lambda=1$ consistently observed across all viscosity ratios. This mode represents the mean flow of the slice-averaged S$_{nw}$ field during steady-state experiments. The mean flow behavior is qualitatively similar to the most dominant structure associated with the first dynamic mode observed in the experimental data of helium jet, as explained in \citep{schmid2011applications}. Henceforth, we consider mode 0 as the most dominant structure and the higher DMD modes capture deviations from this stable trend.


The higher modes, represented by eigenvalues exhibiting harmonic content in Figures \ref{DMDcomponents}(g)-(i), develop in response to the increasing influence of small-scale coherent features in the data. In the M = 0.94 case, there are no other modes present after truncation. However, as M decreases to 0.69, instabilities in flow pathways grow across poorly connected pore spaces in different parts of the micro-core. Thus, the higher modes, such as modes 1 and 3 in Figure \ref{DMDcomponents}(b), begin to display significant perturbations in S$_{nw}$. These signal fluctuations can be attributed to changes in saturation caused by intermittent fluctuations in the occupancy of the nonwetting phase at critical pore locations in the flow pathways across the sample. The temporal dynamics associated with mode 1, i.e., dynamics 1 in Figure \ref{DMDcomponents}(e), exhibits an incomplete sinusoidal cycle. This suggests the presence of long-term transient flow effects in the system, occurring at a smaller spatial scale but on a temporal scale greater than the period of data acquisition used in the experimental study. Conversely, dynamics 3 in Figure \ref{DMDcomponents}(e) displays a slightly damped behavior with a period of approximately 58 seconds, corresponding to mode 3. 

 
Finally, for M = 0.48, the DMD analysis reveals the emergence of more higher modes, as shown in Figure \ref{DMDcomponents}(c). The higher modes, such as modes 2 and 4, are linked to low-energy dominant structures in the data that exhibit relatively high-frequency content. This is evident in their corresponding dynamics 2 and 4, respectively, as shown in Figure \ref{DMDcomponents}(f), when compared to the previous experiment. There is a noticeable increase in the amplitude of these DMD modes near the middle portion of the sample. Additionally, the spatial structures of modes 2 and 4 appear to be similar. In fact, dynamics 4 appears to be the second harmonic of dynamics 2, as evident from the plot in Figure \ref{DMDcomponents}(f), where dynamics 2 and 4 have time periods of approximately 64 seconds and 33 seconds, respectively. In this particular experiment, there is a manifestation of increased amplitude in higher modes near the middle region of the sample. The higher modes, particularly mode 4, exhibit nearly undamped sinusoidal characteristics. 

We observe non-local dynamical features in the spatio-temporal data, alongside the relatively stable saturation field resulting from connected flow pathways across the sample. These low-energy fluctuating features display periodic variations during steady-state experiments.


\begin{algorithm}[t]
\caption{Identification of intermittent behavior using the standard DMD method}
\begin{algorithmic}[1]

\State Prepare the matrix of independent snapshots \{$\mathbf{x_1,...,x_m}$\} based on 2D slice-averaged saturation data.
\State Set \textbf{X} = \{$\mathbf{x_1,...,x_{m-1}}$\} and \textbf{Y} = \{$\mathbf{x_2,...,x_{m}}$\}.
\State Compute the SVD of \textbf{X}, \textbf{X} = \textbf{U} $\mathbf{\Sigma}$ $\mathbf{V^T}$.
\State Define the truncation rank $r$.
\State Compute $\tilde A := \mathbf{U}^T_r$ $\mathbf{Y}$ $\mathbf{V}_r$ $\mathbf{\Sigma}_r^{-1}$. 
\State Compute eigenvalues and eigenvectors of $\tilde{A}\,\mathbf{W}= \boldsymbol{\Lambda} \mathbf{W}$.
\State Compute the amplitudes $b_j$'s of DMD modes.
\State Reconstruct approximated snapshot $x_k$, $x_{k} = \Sigma_{j=1}^{r} \phi_j \lambda_j^{k-1} b_j$.
\State Select DMD modes $\phi_j$ with \{$\mathrm{Im}(\lambda_j) \neq 0 \,\, \mathbf{AND} \,\, 0.99 \leq ||\lambda_j||_2 \leq 1.00$\}.
\State Separate modes whose time dynamics exhibit at least one complete periodic cycle from those with incomplete cycles.


\end{algorithmic}\label{algo1}
\end{algorithm}

\subsection{Identification of intermittent behavior through coherent modes}

In this section, we introduce a workflow to identify regions in the sample exhibiting intermittent behavior by analyzing perturbations in the dynamic modes. It is described as follows:

\begin{enumerate}
    \item Prepare a matrix of independent time snapshots {$\{\mathbf{x_1,...,x_m}\}$\ based on 2D slice-averaged saturation data. 
    \item Set two matrices, \textbf{X} and \textbf{Y}, as \textbf{X} = \{$\mathbf{x_1,...,x_{m-1}}$\} and \textbf{Y} = \{$\mathbf{x_2,...,x_{m}}$\}}. Compute the SVD of matrix \textbf{X}, where \textbf{X} = \textbf{U} $\mathbf{\Sigma}$ $\mathbf{V^T}$.
    \item Identify the truncation rank $r$, and determine $\tilde A := \mathbf{U}^T_r$ $\mathbf{Y}$ $\mathbf{V}_r$ $\mathbf{\Sigma}_r^{-1}$.
    \item Compute the eigenspectrum of reduced rank matrix $\tilde A$, and further calculate the amplitudes of dynamic modes.
    \item Reconstruct the approximated time snapshot $x_k$, $x_{k} = \Sigma_{j=1}^{r} \phi_j \lambda_j^{k-1} b_j$.
    \item DMD modes $\phi_j$ with eigenvalues $\lambda_j$ satisfying the criteria: (1) $ ||\lambda_j||_2 = 1.00$ and (2) $\mathrm{Im}(\lambda_j) = 0$, indicate stable and non-fluctuating behavior of the system. The time dynamics associated with mode 0 were observed to follow these criteria across all experiments.
    \item To identify fluctuating behavior, first select DMD modes $\phi_j$ with $\mathrm{Im}(\lambda_j) \neq 0$. Further, to choose modes reflecting sustained fluctuations, we apply the criterion: $ 0.99 \leq ||\lambda_j||_2 \leq 1.00$. This criterion is designed to preserve coherent structures that consistently exhibit periodic behavior with minimal loss of amplitude in time dynamics. We classify the eigenvalues satisfying this criterion as periodic solutions, represented by star symbols in Figures \ref{DMDcomponents}(g)-(i). The limits of the criterion can be relaxed to identify modes that indicate transient features decaying rapidly, which can be particularly useful in the context of unsteady-state flows.
    \item Finally, distinguish modes whose temporal dynamics exhibit at least one complete cycle from those with incomplete cycles. In the context of experimental data, i.e., 3D snapshots of ganglia, a complete cycle represents the cyclic interplay between drainage and imbibition processes within a given pore space. This includes sequences such as drainage followed by imbibition and then drainage again, or imbibition followed by drainage and then imbibition again, within a local pore space. Time dynamics with one or more complete cycles reflect the cyclic occupancy of fluid, as illustrated later in Sections \ref{m048mode4} and \ref{m069mode3}.
    
\end{enumerate}

\begin{figure}[!t]
\centering
\includegraphics[width=15.5cm]{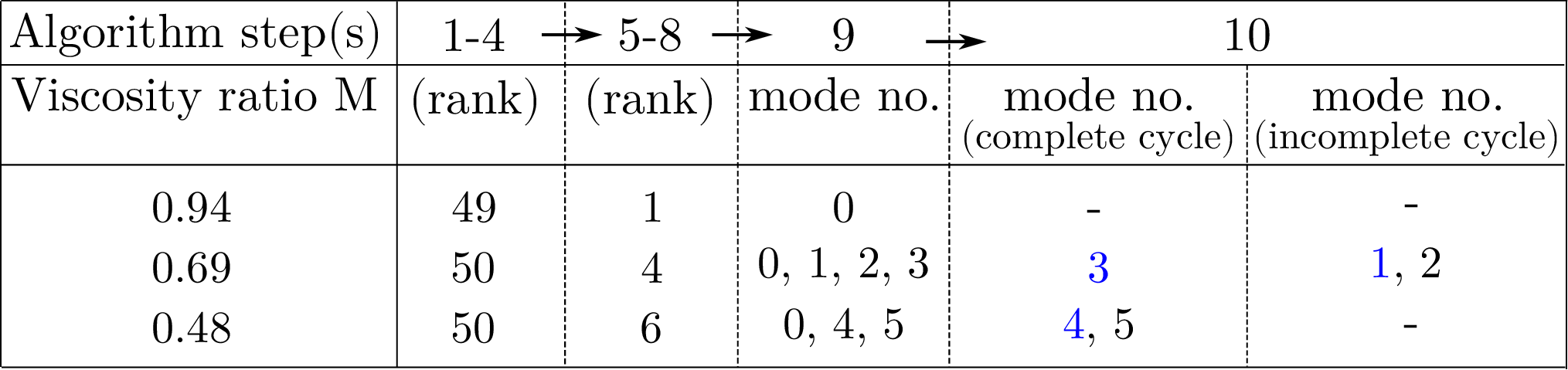}
\caption{The evolution of ranks and identification of modes at different steps of Algorithm \ref{algo1}. (At step 10 of the algorithm, the blue numbers denote the modes selected for later discussion.)}
\label{algorithm}
\end{figure}

The above-described workflow is presented in Algorithm \ref{algo1}. The implementation of the algorithm narrows the spectrum of key coherent structures in the saturation data to five specific dynamic modes for further investigation. The selected modes, at the end of step 10 of the algorithm, are displayed in Figure \ref{algorithm}. Among these, the modes corresponding to dynamics reflecting one or more complete cycles are: (1) modes 4 and 5 in the M = 0.48 case, and (2) mode 3 in the M = 0.69 experiment. In contrast, incomplete cycles are reflected in dynamics 1 and 2, and thus modes 1 and 2 are selected for the M = 0.69 experiment. We will now explore these modes and illustrate the physical insights they provide. For brevity, only one mode of each complex conjugate pair, shown in blue in Figure \ref{algorithm}, is discussed.

\begin{sidewaysfigure}[]
  \centering
  \includegraphics[width=18.5cm]{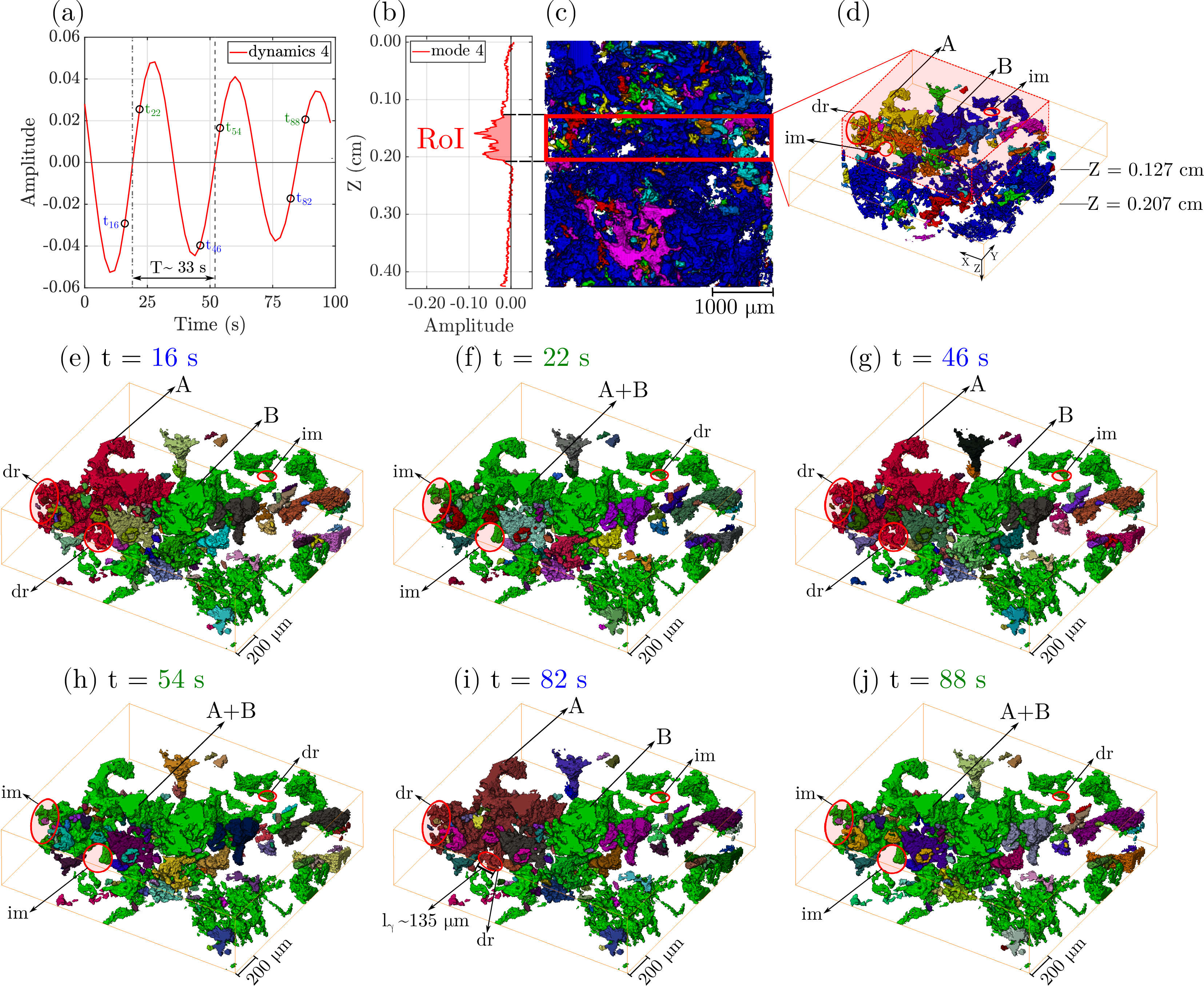}
  \caption{The DMD characteristics are shown for the M = 0.48 case. (a) Time dynamics 4, where circles denote the times at which images of the subvolume are shown in figures (e)-(j); (b) mode 4, where RoI denotes the region of interest; (c) 3D image of the nonwetting phase, where the red rectangle indicates the region of investigation corresponding to the RoI in (b); (d) 3D view of the subvolume extracted from the sample between the specified depths in (c). In (d), the red-shaded cuboid denotes the local region where most fluctuations were observed. A and B represent two clusters of the nonwetting phase, while \textit{dr} and \textit{im} indicate drainage and imbibition processes, respectively, occurring in the pore space highlighted by red circles. The evolution of nonlinear dynamics is illustrated at the time steps: (e) t = 16 s, (f) t = 22 s, (g) t = 46 s, (h) t = 54 s, (i) t = 82 s, and (j) t = 88 s. In (i), the nonwetting fluid drains into the pore space up to a depth of approximately 135 \textmu m. The fluid flow occurs from the bottom toward the top of the sample.}
  \label{mode4m048}
\end{sidewaysfigure}

\subsubsection{Mode 4 in the M = 0.48 experiment}\label{m048mode4}

The time dynamics associated with mode 4 is shown in Figure \ref{mode4m048}(a). Based on the total time of collected scans, which is 98 seconds, we anticipate observing approximately 3 cycles of intermittent events, given the average time period T of 32.6 seconds between cycles. The spatial structure of this mode, displayed in Figure \ref{mode4m048}(b), reveals the presence of both small and large-scale fluctuations along the length of the sample. To illustrate the physical behavior of the system captured by this mode, we investigate the region responsible for generating large-scale fluctuations. This is indicated by the region of interest (RoI) shaded in red in Figure \ref{mode4m048}(b). The corresponding depth in the sample, ranging from Z = 0.127 cm to 0.207 cm, is marked by the red rectangle in Figure \ref{mode4m048}(c). Figure \ref{mode4m048}(d) displays the subvolume extracted from the sample within the specified depths or RoI. After extracting and examining the sequence of images, significant topological changes in the nonwetting phase occupancy were observed in this subvolume. These changes most notably occurred within the region highlighted by the red-shaded cuboid, as shown in Figure \ref{mode4m048}(d). In Figure \ref{mode4m048}(d), A and B represent two clusters of the nonwetting phase, while \textit{im} and \textit{dr} denote the physical processes of imbibition and drainage, respectively, occurring within the localized pore spaces highlighted by red circles. Figures \ref{mode4m048}(e)-(j) illustrate the periodic changes in the nonwetting phase occupancy observed at different time steps within the subvolume highlighted in Figure \ref{mode4m048}(d). It is important to note that cluster B is connected to a major ganglion, which is observed to extend from the inlet to the outlet of the sample. The red circles labeled \textit{dr} in Figures \ref{mode4m048}(e), \ref{mode4m048}(g), and \ref{mode4m048}(i) indicate the evolution of drainage at different time steps, which result in the growth of cluster A through an increase in the volume of the nonwetting phase, or in other words, an increase in S$_{nw}$. Simultaneously, the connected flow pathway through cluster B contains brine in the pore space highlighted by circle \textit{im}. On the other hand, in Figures \ref{mode4m048}(f), \ref{mode4m048}(h) and \ref{mode4m048}(j), cluster A shows relaxation dynamics with imbibition occurring in the same pore spaces that were previously drained by the nonwetting phase. In these concentrated spaces, the external flow energy accumulated during drainage dissipated, causing menisci retraction as brine invaded several pores sequentially in a pore-filling event. Interestingly, during these periods, clusters A and B coalesce, as illustrated by the single cluster A+B (in green color) in Figures \ref{mode4m048}(f), \ref{mode4m048}(h), and \ref{mode4m048}(j). A significant portion of the large accumulated nonwetting flux is transported downstream through drainage across the connected clusters. This is evidenced by an increase in nonwetting phase volume in the same pore space that was previously imbibed with brine in cluster B (denoted by \textit{im} in Figures \ref{mode4m048}(e), \ref{mode4m048}(g) and \ref{mode4m048}(i)).

The volume of fluid $V_{f}^k$ in the RoI intermittently occupying the pore space at time snapshot $k$ for a particular mode can be estimated using the following expression:
\begin{equation} \label{eq7}
{V_f}^k = \Sigma_{i=m}^{n}{\mathrm{S}_{nw}}^k_{i}\,{{V}_{p}}_i \,,  
\end{equation}
where the indices $m$ and $n$ represent the start and end slice numbers, respectively, corresponding to the RoI. The term ${\mathrm{S}_{nw}}^k_{i}$ denotes the saturation of the nonwetting phase in slice $i$ at time snapshot $k$ for the given mode, and ${{V}_{p}}_i$ represents the volume of pores in slice $i$. Here, \({\mathrm{S}_{nw}}^k_{i}\) can be calculated using Equation \ref{eq3} (considering complex conjugate pairs of modes) and varies for different modes within the same RoI. At t = 82 seconds, $V^{82\, \mathrm{s}}_f$ for mode 4 between depths at Z = 0.127 cm and 0.207 cm is calculated to be $1.86 \times 10^{6}$ \textmu m$^3$. This fluctuating volume of fluid ($V_{f}^k \sim l_\gamma^3$) provides a characteristic perturbation length $l_\gamma$ of 123 \textmu m at the pore-scale. Based on experimental observations, the nonwetting phase drains into the pores near the bottom of the cuboid, as indicated by \textit{dr} in Figure \ref{mode4m048}(i), extending up to a length of approximately 135 \textmu m. This is in close agreement with the perturbation length estimated using DMD.

In summary, in the absence of cluster migration, the interrupted transport of the nonwetting phase is observed to occur through intermittently connecting ganglia in the sample. The dynamic mode reflected high-amplitude signals corresponding to these fluctuations in the saturation field where significant cyclic occupancy of different fluids could be observed through the complementary transport processes of drainage and imbibition. Interestingly, the RoI also displays the manifestation of non-local dynamics in different regions of the sample. This is reflected in the synchronous occurrence of drainage and imbibition processes across distant pores. In addition, the sequence of major drainage events shown in Figures \ref{mode4m048}(e), \ref{mode4m048}(g), and \ref{mode4m048}(i), and the major imbibition events displayed in Figures \ref{mode4m048}(f), \ref{mode4m048}(h), and \ref{mode4m048}(j) share highly similar topological configurations. The average time difference between consecutive occurrences of drainage (and similarly, imbibition) events is approximately 33 seconds, in accordance with the time dynamics corresponding to mode 4. Similar observations were noted in other parts of the sample for small-scale fluctuations indicated by mode 4, albeit with reduced accuracy.



\begin{figure}[!t]
\centering
\includegraphics[width=16cm]{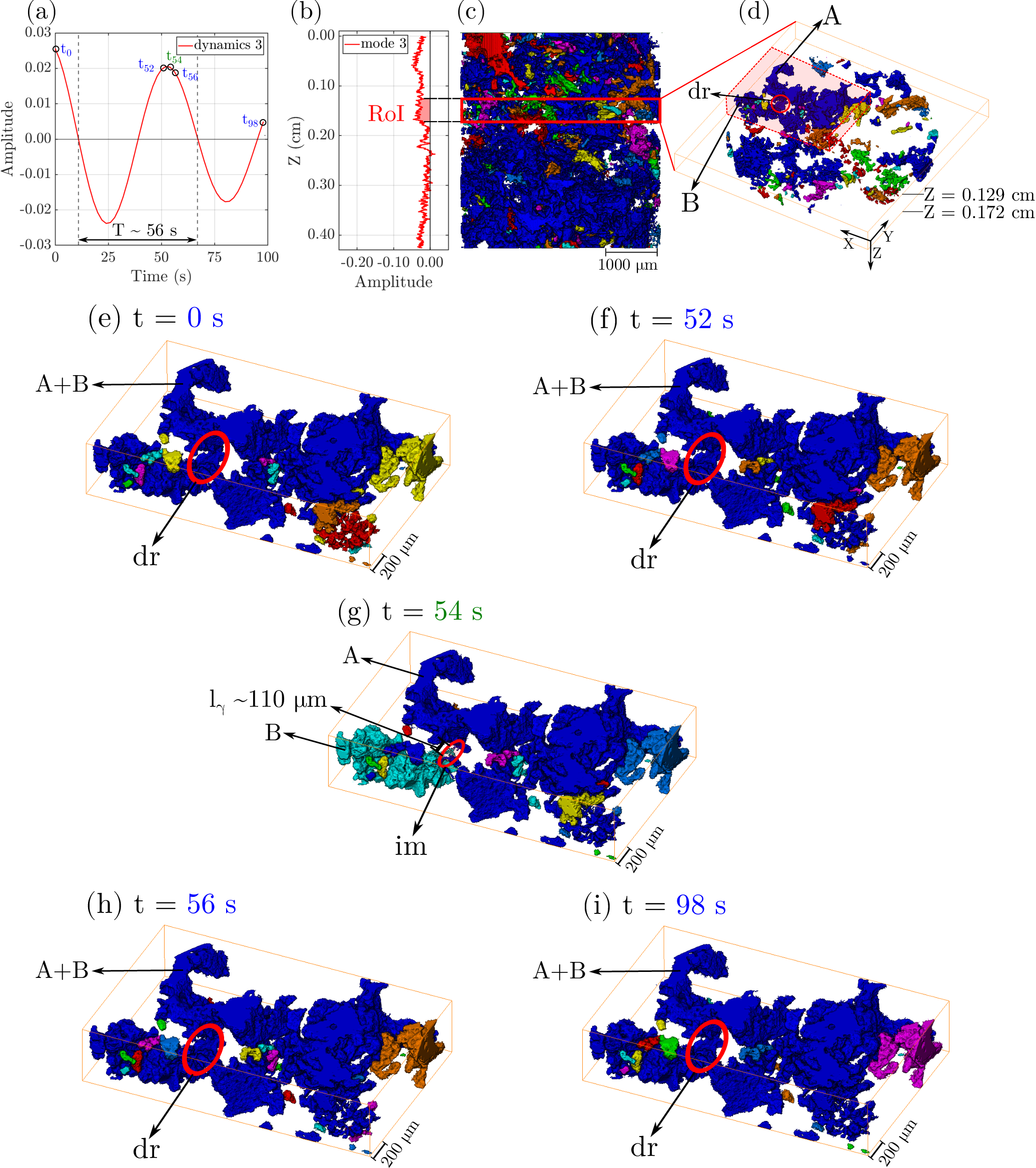}
\caption{The DMD characteristics are shown for the M = 0.69 case. (a) Time dynamics 3, where circles denote the times at which images of the subvolume are shown in figures (e)-(i); (b) mode 3, where RoI denotes the region of interest; (c) 3D image of the nonwetting phase, where the red rectangle indicates the region of investigation corresponding to the RoI in (b); (d) 3D view of the subvolume extracted from the sample between the specified depths in (c). In (d), the red-shaded cuboid denotes the local region where significant fluctuations were observed. A and B represent two clusters of the nonwetting phase, while \textit{dr} indicates the drainage processes occurring in the pore space highlighted by the red circle. The evolution of nonlinear dynamics is illustrated at the time steps: (e) t = 0 s, (f) t = 52 s, (g) t = 54 s, (h) t = 56 s, and (i) t = 98 s. In (g), the nonwetting fluid drains into the pore space up to a depth of approximately 110 \textmu m. The fluid flow occurs from the bottom toward the top of the sample.}
\label{mode3m69}
\end{figure}

\subsubsection{Mode 3 in the M = 0.69 experiment}\label{m069mode3}

The time dynamics associated with mode 3 is shown in Figure \ref{mode3m69}(a). In this case, based on the total time of 98 seconds, we expect to observe 1 complete cycle of an intermittent event, given the time period T of 56 seconds. The dynamic mode shown in Figure \ref{mode3m69}(b) exhibits both small and large-scale fluctuations along the length of the sample. Large amplitude fluctuations are observed near the mid-section of the sample, a region previously inferred to be highly heterogeneous. We investigate a subvolume that reflects large-scale perturbations, indicated by RoI and shaded in red in Figure \ref{mode3m69}(b). The corresponding depth in the sample, ranging from Z = 0.129 cm to 0.172 cm, is marked by the red rectangle in Figure \ref{mode3m69}(c). Figure \ref{mode3m69}(d) shows the subvolume extracted from the sample based on the RoI. Furthermore, the red-shaded cuboid in the extracted subvolume highlights the local area where significant menisci rearrangements were observed during steady-state flow. In Figure \ref{mode3m69}(d), A and B represent two clusters of the nonwetting phase, which are connected through local pore space drained by the nonwetting phase, indicated by the red circle labeled \textit{dr}. The time steps shown in Figures \ref{mode3m69}(e)-(i) illustrate the occurrence of an intermittent event in the subvolume highlighted in Figure \ref{mode3m69}(d). During the time period between t = 0 seconds and 52 seconds, as shown in Figures \ref{mode3m69}(e) and \ref{mode3m69}(f), respectively, clusters A and B remain connected. This is depicted as cluster A+B in dark blue, which is part of the major nonwetting ganglion extending from the inlet to the outlet of the sample. The red circle labeled \textit{dr} in Figures \ref{mode3m69}(e) and \ref{mode3m69}(f) shows the local pore space continuously drained by the nonwetting phase. However, at t = 54 seconds, as shown in Figure \ref{mode3m69}(g), the pore space is imbibed with brine, indicated by the circle marked \textit{im}. This results in the fragmentation of cluster A+B into distinct clusters A and B (depicted in different colors), with A remaining connected to the major ganglion. Subsequently, at t = 56 seconds, fluid rearrangement occurs, returning the configuration to nearly the same arrangement observed at t = 52 seconds, as shown in Figure \ref{mode3m69}(f). The two clusters, A and B, coalesce, with the interconnecting pore space once again drained by the nonwetting phase, as shown in Figure \ref{mode3m69}(h). This completes a sequence of events related to the cyclic occupancy of two fluids in a given pore space. Finally, the existing saturation field configuration remains almost unchanged until the final observation at t = 98 seconds, as depicted in Figure \ref{mode3m69}(i). 

At t = 54 seconds, the fluctuating volume of fluid is calculated for mode 3 between depths at Z = 0.129 cm and 0.172 cm using Equation \ref{eq7}. $V^{54\, \mathrm{s}}_f$ is evaluated to be $1.23 \times 10^{6}$ \textmu m$^3$. This fluctuating volume of fluid provides a characteristic perturbation length $l_\gamma$ of 107 \textmu m at the pore-scale. The wetting phase imbibes in the pores near the middle of the cuboid, as indicated by \textit{im} in Figure \ref{mode3m69}(g), up to a depth of approximately 110 \textmu m. This is in excellent agreement with the perturbation length calculated using DMD.

The time period corresponding to dynamics 3, which is 56 seconds, has been accurately validated through the analysis of 4D images of the experimental data. Similar to the M = 0.48 experiment, the aforementioned sequence of events provides evidence of low-energy transients occurring in the multi-phase flow field under steady-state conditions. Such transient effects can potentially increase temporal variability in the relative motion between the interacting phases, thereby impacting relative permeabilities at fixed volumetric saturation. It should be noted that phase topology and distribution of pore sizes in porous media are complex. As a result, pore-filled connections between clusters of different phases can range from a few to a large number of voxels, depending on the intricate geometry of the porous microstructure. Therefore, the resolution of voxels and the methods used for image segmentation are important in generating an accurate topological realization of the saturation field. In this respect, the proposed DMD analysis using slice-averaged saturation values may identify dynamic behavior in small fluid-filled pore spaces (represented by a few voxels) through higher modes, albeit with some ambiguity and reduced accuracy. In such cases, applying DMD within a 3D spatial field can yield reliable and more accurate screening results by effectively tracking small-scale saturation fluctuations within the sample space.

\subsubsection{Mode 1 in the M = 0.69 experiment}

\begin{figure}[!t]
\centering
\includegraphics[width=16.5cm]{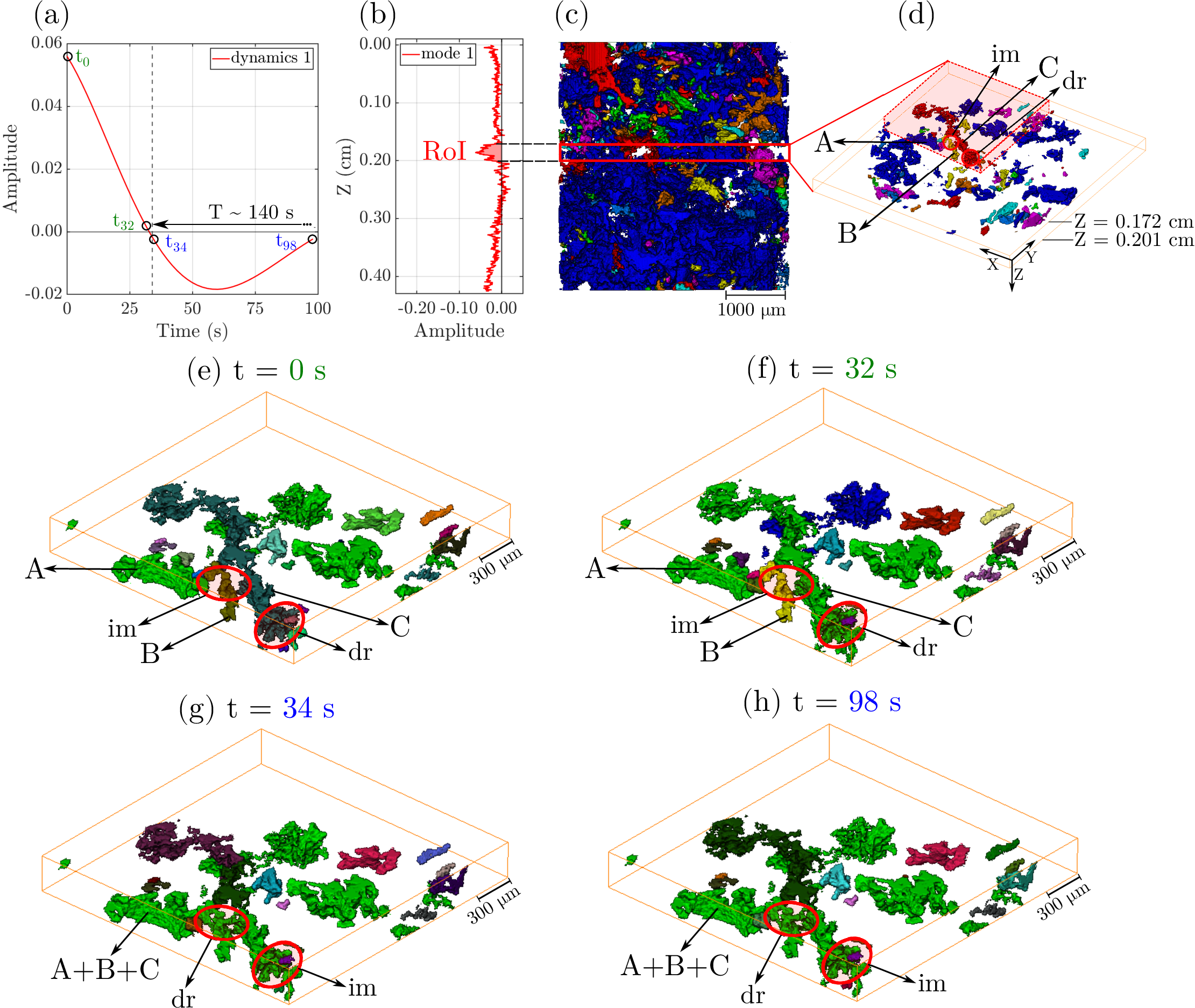}
\caption{The DMD characteristics are shown for the M = 0.69 case. (a) Time dynamics 1, where circles denote the times at which images of the subvolume are shown in figures (e)-(h); (b) mode 1, where RoI denotes the region of interest; (c) 3D image of the nonwetting phase, where the red rectangle indicates the region of investigation corresponding to the RoI in (b); (d) 3D view of the subvolume extracted from the sample between the specified depths in (c). In (d), the red-shaded cuboid denotes the local region where most fluctuations were observed. A, B and C, represent three clusters of the nonwetting phase, while \textit{im} indicates the imbibition processes occurring in the pore space highlighted by the red circle. The evolution of nonlinear dynamics is illustrated at the time steps: (e) t = 0 s, (f) t = 32 s, (g) t = 34 s, and (h) t = 98 s. The fluid flow occurs from the bottom toward the top of the sample.}
\label{mode1m69}
\end{figure}

In DMD, it is generally recommended to include an entire cycle within the historical time snapshots. This step is recommended because the method relies on modal decomposition and frequency analysis, which are particularly suitable for analyzing dynamic systems with periodic behavior \citep{li2024prediction}. Nevertheless, using historical time steps that do not cover a complete cycle can reveal potential long-term dynamics in multi-phase flows in porous media. Experimental studies have shown that relaxation times for complex fluid-fluid displacements can range from seconds \citep{armstrong2014subsecond} to hours \citep{schluter2017time}. In this section, we analyze an instance of prolonged dynamics in the M = 0.69 experiment. This scenario is illustrated through mode 1 following the application of the proposed workflow.


The time dynamics associated with mode 1 is shown in Figure \ref{mode1m69}(a). The oscillation frequency $f_i$ corresponding to an eigenvalue $\lambda_i$ of a mode $\phi_i$ is determined through \(f_i=\nicefrac{\angle\lambda_i}{(2 \pi\Delta t_{scan})}\), with $\Delta t_{scan}$ denoting the uniform sampling increment between snapshot pairs. The time period $\mathrm{T}$ ($= \nicefrac{1}{f}$) of the cycle for dynamics 1 is estimated to be approximately 140 seconds. This indicates an incomplete cycle, as the total duration of the scans is only 98 seconds. The dynamic mode 1 is shown in Figure \ref{mode1m69}(b). We investigate a subvolume near the mid-section of the sample, indicated as RoI in Figure \ref{mode1m69}(b). The corresponding depth in the sample, ranging from Z = 0.172 cm to 0.201 cm, is marked by the red rectangle in Figure \ref{mode1m69}(c). Figure \ref{mode1m69}(d) displays the subvolume extracted from the sample between the specified depths. The red-shaded cuboid in Figure \ref{mode1m69}(d) highlights the local region where potential long-term dynamics could be observed during steady-state flow. In Figure \ref{mode1m69}(d), A, B, and C represent three clusters of the nonwetting phase. Here, \textit{im} (or \textit{dr}) denotes the imbibition (or drainage) process occurring within the local pore spaces highlighted by red circles. For the time interval between t = 0 seconds and 32 seconds, as shown in Figures \ref{mode1m69}(e) and \ref{mode1m69}(f), respectively, clusters A, B, and C remain disconnected. The interconnecting pore spaces between the clusters contain brine and are labeled as \textit{im} in the figures. Also, a large flux of the nonwetting phase is accumulated in cluster C, in the region denoted by \textit{dr} in Figures \ref{mode1m69}(e) and \ref{mode1m69}(f). At t = 34 seconds, the three clusters are observed to coalesce, forming a single entity labeled as cluster A+B+C, depicted in green in Figure \ref{mode1m69}(g). The individual clusters become connected through isthmuses of pore spaces drained by the nonwetting phase, denoted as \textit{dr} in Figure \ref{mode1m69}(g). This fluid spreading causes non-local dynamic effects on fluid menisci in distant pores. The change is noticeable as weak imbibition occurs in the pore space (denoted as \textit{im} in Figure \ref{mode1m69}(g)), which was previously occupied by the nonwetting phase in cluster C. The existing saturation field configuration remains almost unchanged until the final time at t = 98 seconds, as shown in Figure \ref{mode1m69}(h). 

Thus, in this experiment, mode 1 highlights a key coherent structure that identifies regions exhibiting partial cyclic changes in fluid occupancy, transitioning only from imbibition to drainage, or vice versa. Based on a sparse set of measurements, this DMD analysis predicts that certain transient dynamics either occur at time scales greater than 2 minutes or indicate some aperiodic behavior in the system.


\section{Conclusions}

In this study, we demonstrated the application of Dynamic Mode Decomposition as a diagnostic tool to rapidly identify locations in heterogeneous pore space where large-scale intermittent fluid occupancy was observed during fractional flow experiments under steady-state conditions. We developed a novel workflow that first performs dimensionality reduction of high-dimensional, feature-rich datasets consisting of 4D images. It then effectively scans the dominant coherent structures, represented by dynamic modes and their associated time dynamics, to navigate the spatial regions exhibiting low-energy oscillatory behavior in multi-phase flows. 

Intermittency in fluid occupancy was observed to increase as the viscosity ratio decreased. We estimated that 99\% of the total variance in the saturation datasets could be captured with substantially reduced number of ranks. The data reconstructed using reduced ranks exhibited only marginal error compared to the original datasets. We presented three cases of modes where both short- and long-term dynamics were observed. In the M = 0.48 and M = 0.69 experiments, the short-term dynamics, exhibiting three and one complete cycle respectively, accurately matched the DMD results. In addition, the non-intrusive DMD technique inferred potential long-term dynamics in the M = 0.69 case from a sparse set of measurements through the presence of an incomplete cycle in the time dynamics. By directly scanning the local subvolume in the sample, identified through the mode, across the time-series of 3D images, an incomplete sequence of intermittent events was observed in this case. This is particularly significant because the duration of high-resolution synchrotron tomography imaging is limited due to its high costs. Such data-driven insights can provide valuable information about the long-term behavior of the system and assist in determining an appropriate Representative Elementary Volume (REV) size for upscaling flow parameters from the pore scale to the continuum scale \citep{wang2023pore}.

We demonstrate that DMD is highly effective at capturing complex nonlinear behaviors and modal interactions. The modes and their associated dynamics revealed extensive insights into the flow dynamics. Notably, the high-amplitude signals observed in the modes corresponded to non-local dynamics spanning distant pores. Further investigation within the subvolume revealed that these non-local dynamics arose from complex fluid displacement patterns, including the fragmentation and coalescence of clusters. We proposed a new analytical approach for estimating the fluid volume fluctuating at different frequencies within a given pore space, based on mode-dynamic pairs. The characteristic length scale of perturbation in the saturation field was validated with high accuracy against experimental observations.

The proposed algorithm, being computationally inexpensive and fast, can be readily integrated into the pre-processing stages of advanced characterization methods developed for feature-rich experimental datasets, such as those in \citep{reynolds2017dynamic,spurin2019intermittent,bultreys20244d}.  Future applications of DMD and its variants, such as multi-resolution DMD \citep{kutz2016multiresolution}, in a complete 3D saturation field configuration could be explored to further accelerate the identification of critical pore spaces that control intermittency. Visualization of dominance structures, as discussed by \citep{krake2021visualization}, can help in understanding the evolving significance of modes over time. In addition, integrating DMD with machine learning approaches could facilitate robust analysis of 4D imaging data across diverse rock types, different pairs of resident/injected fluids, and flow regimes. This could enable the development of a library of phenomenological relationships pertinent to intermittent flow behavior, which would be useful for accurately upscaling flow parameters while accounting for pore-scale dynamics.


\section*{Acknowledgements}
\noindent A. Raizada, S. M. Benson, H. A. Tchelepi, and C. Spurin acknowledge the support from the GeoCquest consortium.



\bibliographystyle{elsarticle-harv} 
\bibliography{references}





\end{document}